\documentclass[useAMS,usenatbib]{mn2e}

\usepackage{graphicx,amssymb,amstext} 

\begin{document}

\title[Detection of cosmic magnification of sub-mm galaxies]{HerMES: detection of cosmic magnification of sub-mm galaxies using angular cross-correlation\thanks{{\it Herschel} is an ESA space observatory with science instruments provided by European-led Principle Investigator consortia and with important participation from NASA.}}
\author[L.~Wang et al.]
{\parbox{\textwidth}{\raggedright L.~Wang,$^{1}$\thanks{E-mail: \texttt{lingyu.wang@sussex.ac.uk}}
A.~Cooray,$^{2,3}$
D.~Farrah,$^{1}$
A.~Amblard,$^{2}$
R.~Auld,$^{4}$
J.~Bock,$^{3,5}$
D.~Brisbin,$^{6}$
D.~Burgarella,$^{7}$
P.~Chanial,$^{8}$
D.L.~Clements,$^{9}$
S.~Eales,$^{4}$
A.~Franceschini,$^{10}$
J.~Glenn,$^{11}$
Y.~Gong, $^{2}$
M.~Griffin,$^{4}$
S.~Heinis,$^{7}$
E.~Ibar,$^{12}$
R.J.~Ivison,$^{12,13}$
A.M.J.~Mortier,$^{9}$
S.J.~Oliver,$^{1}$
M.J.~Page,$^{14}$
A.~Papageorgiou,$^{4}$
C.P.~Pearson,$^{15,16}$
I.~P{\'e}rez-Fournon,$^{17,18}$
M.~Pohlen,$^{4}$
J.I.~Rawlings,$^{14}$
G.~Raymond,$^{4}$
G.~Rodighiero,$^{10}$
I.G.~Roseboom,$^{1}$
M.~Rowan-Robinson,$^{9}$
Douglas~Scott,$^{19}$
P.~Serra,$^{2}$
N.~Seymour,$^{14}$
A.J.~Smith,$^{1}$
M.~Symeonidis,$^{14}$
K.E.~Tugwell,$^{14}$
M.~Vaccari,$^{10}$
J.D.~Vieira,$^{3}$
L.~Vigroux$^{20}$ and
G.~Wright$^{12}$}\vspace{0.4cm}\\
\parbox{\textwidth}{\raggedright $^{1}$Astronomy Centre, Dept. of Physics \& Astronomy, University of Sussex, Brighton BN1 9QH, UK\\
$^{2}$Dept. of Physics \& Astronomy, University of California, Irvine, CA 92697, USA\\
$^{3}$California Institute of Technology, 1200 E. California Blvd., Pasadena, CA 91125, USA\\
$^{4}$Cardiff School of Physics and Astronomy, Cardiff University, Queens Buildings, The Parade, Cardiff CF24 3AA, UK\\
$^{5}$Jet Propulsion Laboratory, 4800 Oak Grove Drive, Pasadena, CA 91109, USA\\
$^{6}$Space Science Building, Cornell University, Ithaca, NY, 14853-6801, USA\\
$^{7}$Laboratoire d'Astrophysique de Marseille, OAMP, Universit\'e Aix-marseille, CNRS, 38 rue Fr\'ed\'eric Joliot-Curie, 13388 Marseille cedex 13, France\\
$^{8}$Laboratoire AIM-Paris-Saclay, CEA/DSM/Irfu - CNRS - Universit\'e Paris Diderot, CE-Saclay, pt courrier 131, F-91191 Gif-sur-Yvette, France\\
$^{9}$Astrophysics Group, Imperial College London, Blackett Laboratory, Prince Consort Road, London SW7 2AZ, UK\\
$^{10}$Dipartimento di Astronomia, Universit\`{a} di Padova, vicolo Osservatorio, 3, 35122 Padova, Italy\\
$^{11}$Dept. of Astrophysical and Planetary Sciences, CASA 389-UCB, University of Colorado, Boulder, CO 80309, USA\\
$^{12}$UK Astronomy Technology Centre, Royal Observatory, Blackford Hill, Edinburgh EH9 3HJ, UK\\
$^{13}$Institute for Astronomy, University of Edinburgh, Royal Observatory, Blackford Hill, Edinburgh EH9 3HJ, UK\\
$^{14}$Mullard Space Science Laboratory, University College London, Holmbury St. Mary, Dorking, Surrey RH5 6NT, UK\\
$^{15}$Space Science \& Technology Department, Rutherford Appleton Laboratory, Chilton, Didcot, Oxfordshire OX11 0QX, UK\\
$^{16}$Institute for Space Imaging Science, University of Lethbridge, Lethbridge, Alberta, T1K 3M4, Canada\\
$^{17}$Instituto de Astrof{\'\i}sica de Canarias (IAC), E-38200 La Laguna, Tenerife, Spain\\
$^{18}$Departamento de Astrof{\'\i}sica, Universidad de La Laguna (ULL), E-38205 La Laguna, Tenerife, Spain\\
$^{19}$Department of Physics \& Astronomy, University of British Columbia, 6224 Agricultural Road, Vancouver, BC V6T~1Z1, Canada\\
$^{20}$Institut d'Astrophysique de Paris, UMR 7095, CNRS, UPMC Univ. Paris 06, 98bis boulevard Arago, F-75014 Paris, France}}

\date{Accepted . Received ; in original form }

\maketitle

\begin{abstract}
Cosmic magnification is due to the weak gravitational lensing of sources in the distant Universe by foreground large-scale structure leading to coherent changes in the observed number density of the background sources. Depending on the slope of the background source number counts, cosmic magnification causes a correlation between the background and foreground galaxies, which is unexpected in the absence of lensing if the two populations are spatially disjoint. Previous attempts using submillimetre (sub-mm) sources have been hampered by small number statistics. The large number of sources detected in the {\it Herschel} Multi-tiered Extra-galactic Survey (HerMES) Lockman-SWIRE field enables us to carry out the first robust study of the cross-correlation between sub-mm sources and sources at lower redshifts. Using ancillary data we compile two low-redshift samples from SDSS and SWIRE with $\langle z\rangle \sim 0.2$ and 0.4, respectively, and cross-correlate with two sub-mm samples based on flux density and colour criteria, selecting galaxies preferentially at $z \sim 2$. We detect cross-correlation on angular scales between $\sim1$ and 50 arcmin and find clear evidence that this is primarily due to cosmic magnification. A small, but non-negligible signal from intrinsic clustering is likely to be present due to the tails of the redshift distribution of the sub-mm sources overlapping with those of the foreground samples. 
\end{abstract}

\clearpage


\section{INTRODUCTION}
Large-scale structure at low redshifts systematically magnifies sources at higher redshifts as a result of gravitational light deflection in the weak limit. On the one hand, fewer sources will be observed, because lensing stretches the solid angle and dilutes the surface density of sources. Conversely, the effective flux limit is lowered as a result of magnification, which leads to a deeper survey. Whether there is an increase or decrease in the observed number density of sources depends on the shape of the background source number counts -- an effect known as the magnification bias (Bartelmann \& Schneider 2001; hereafter BS01). At submillimetre (sub-mm) wavelengths the magnification bias is expected to be large and positive, resulting in an increase in the observed number density of sources compared to the case without lensing (e.g. Blain \& Longair 1993; Blain et al. 1996; Negrello et al. 2007; Lima et al 2010).

Cosmic magnification also induces an apparent angular cross-correlation between two source populations with disjoint spatial distributions. It can thus be measured by cross-correlating non-overlapping foreground and background samples. When combined with number counts, such a cross-correlation can provide constraints on cosmological parameters (e.g. $\Omega_{{\rm m}}, \sigma_8$) and galaxy bias, a key ingredient in galaxy formation and evolution models (M\'{e}nard \& Bartelmann 2002). As the weak lensing-induced cross-correlation also probes the dark matter distribution, it provides an independent cross-check of the cosmic shear measurements, which depend on the fundamental assumption that galaxy ellipticities are intrinsically uncorrelated. Most previous investigations, using foreground galaxies selected in the optical or infrared together with background quasars, have produced controversial or inconclusive results (e.g. Seldner \& Peebles 1979; Bartelmann \& Schneider 1994; Bartsch et al. 1997). The best detection to date is presented in Scranton et al. (2005), where cosmic magnification is detected at an $8\sigma$ significance level using 13 million galaxies and $\sim$200,000 quasars from the Sloan Digital Sky Survey (SDSS).

The amplitude of the weak lensing-induced cross-correlation is determined by several factors: the dark matter power spectrum and growth function, the shape of the background source number counts and the bias of the foreground sources. At sub-mm wavelengths, the power-law slope of the cumulative number count is exceptionally steep, $>$2.5 for sources in the flux range $0.02-0.5$ Jy at 250, 350 and 500 $\mu$m (e.g. Patanchon et al. 2009; Oliver et al. 2010a; Glenn et al. 2010; Clements et al. 2010). In Scranton et al. (2005), the number count slope of the quasar sample is considerably flatter ($\sim2$ for the brightest ones). In addition, sub-mm sources detected in deep surveys mainly reside in the high-redshift Universe with a median redshift of $z\sim$ 2 (Chapman et al. 2003, 2005; Pope et al. 2006; Aretxaga et al. 2007; Amblard et al. 2010). The steep number counts, together with the large redshift range, make sub-mm sources an ideal background sample. So far there have been two attempts at measuring the weak lensing-induced cross-correlation between foreground optical galaxies and background sub-mm sources, but with conflicting results. Almaini et al. (2005) measured the cross-correlation between 39 SCUBA sources and optical sources at lower redshifts $\langle z \rangle \sim 0.5$. They claimed evidence for a significant signal which might be caused by lensing. Conversely, Blake et al. (2006) did not find evidence for cross-correlation due to cosmic magnification using a similar number of sources.

The {\it Herschel} Multi-tiered Extra-galactic Survey (HerMES, Oliver et al. 2010b) is the largest project being undertaken by {\it Herschel} (Pilbratt et al. 2010). In this paper, we calculate the angular cross-correlation between foreground galaxies selected from SDSS or the {\it Spitzer} Wide-area Infrared Extragalactic (SWIRE; Lonsdale 2003, 2004) survey and background sub-mm sources detected by the Spectral and Photometric Imaging Receiver (SPIRE; Griffin et al. 2010) instrument on {\it Herschel}. This paper is organised as follows: In Section 2, we give a brief introduction to magnification bias and the angular cross-correlation function. In Section 3, we describe the various data-sets used as foreground and background samples. Measurements of the cross-correlation between foreground and background samples are presented in Section 4. Finally, discussions and conclusions are given in Section 5. Throughout the paper, we use a spatially flat $\Lambda$CDM cosmology with $\Omega_{{\rm m}} = 0.3$ and $H_0=70$ km s$^{-1}$ Mpc$^{-1}$. Magnitudes are in the AB system.

\section{Modelling the cross-correlation function}

 In this section we briefly describe the magnification bias and how it manifests itself in the number density and cross-correlation between two spatially separated populations. We refer the reader to Moessner \& Jain (1998), BS01, Cooray \& Sheth (2002) and references therein for a complete introduction. Suppose a background population has an intrinsic (i.e. unlensed) number density $n_{{\rm u}}(S, z)$, where $S$ is flux density and $z$ is redshift. As a result of lensing, the sky solid angle is stretched locally by a factor of $\mu(\hat{\phi}, z)$ ($\hat{\phi}$ denotes angular position on the sky) and $S$ is magnified by the same factor because surface brightness is preserved. The two contrasting effects modify the observed (lensed) number density in the following way 
\begin{equation}
n_{{\rm l}}(S, z) = \frac{n_{{\rm u}}(S/\mu(\hat{\phi}, z), z)}{\mu(\hat{\phi}, z)} .
\end{equation}
When the lens plane is at a much lower redshift than the source plane, the redshift-dependent magnification can be substituted by the magnification $\mu$ of a source at infinity. Assuming the cumulative number count distribution of the background population can be described by a power-law $N_{{\rm u}}(S) \propto S^{-\beta}$, we should expect a factor of 
\begin{equation}
\frac{N_{{\rm l}}(S)}{N_{{\rm u}}(S)} = \mu^{\beta - 1}
\end{equation}
 change in the observed number count. Strictly speaking, the number count slope $\beta=\beta(S)$ is a function of flux density. In this paper, we make the simplifying assumption that $\beta$ is a constant over the flux range we probe. Using the number counts of resolved sources presented in Oliver et al. (2010a), we find that in the flux range $0.03 - 0.5$ Jy, $\beta=2.53\pm0.16$, $2.99\pm0.51$ and $2.66\pm0.24$ at 250, 350 and 500 $\mu$m respectively.

The angular cross-correlation function between population 1 at lower redshifts and population 2 at higher redshifts is defined as 
\begin{equation}
w_{\textrm{cross}}(\theta) = \langle \delta n_1(\hat{\phi}) \delta n_2(\hat{\phi'}) \rangle, 
\end{equation}
where $\delta n_i \equiv n_i(\hat{\phi}) / \bar{n}_i - 1$ is the number density fluctuation and $\bar{n}_i$ is the average number density of the $i$th sample. We can decompose $\delta n_i$ into two parts,
\begin{equation}
\delta n_i (\hat{\phi}) = \delta n_i^{{\rm c}} (\hat{\phi}) + \delta n_i^{\mu} (\hat{\phi}).
\end{equation}
The first term $\delta n_i^{{\rm c}}$ is due to intrinsic clustering of galaxies and is a projection of density fluctuations along the line-of-sight, 
\begin{equation}
\delta n_i^{{\rm c}} (\hat{\phi}) = b_i \int_0^{\chi_{{\rm H}}} d \chi W_i(\chi) \delta(r(\chi)\hat{\phi}, a),
\end{equation}
where $\chi_{{\rm H}}$ is the comoving radial distance to the horizon, $r(\chi)$ is the comoving angular diameter distance, $W(\chi)$ is the normalised radial distribution of the sources, $a$ is the scale factor, $\delta(r(\chi)\hat{\phi}, a)$ is the dark matter density perturbations and $b_i$ is the bias factor assumed to be scale- and time-independent. The second term in equation (4) $\delta n_i^{\mu}$ is due to magnification bias,
\begin{equation} 
\delta n_i^{\mu} = \frac{N_{{\rm l}} - N_{{\rm u}}}{N_{{\rm u}}}=  \mu^{\beta - 1} - 1 = 2 (\beta - 1) \kappa. 
\end{equation}
In the last step, we have used the weak lensing limit, $\mu = 1 + 2\kappa$. The convergence $\kappa$ is a weighted projection of the density field along the line of sight (BS01), 
\begin{equation}
\kappa_i(\hat{\phi}) = \frac{3}{2} \Omega_{{\rm m}} \int_0^{\chi_{{\rm H}}} d\chi g_i(\chi) \frac{\delta(r\hat{\phi}, a)}{a},
\end{equation}
where $g(\chi)$ is the radial weighting function defined as 
\begin{equation}
g(\chi) \equiv r(\chi) \int_{\chi}^{\chi_{{\rm H}}} \frac{r(\chi' -\chi)}{r(\chi')} W(\chi')d \chi'.
\end{equation}
The angular cross-correlation between the two populations is then
\begin{eqnarray}
w_{\textrm{cross}}(\theta) & = & \langle \delta n_1^{{\rm c}} (\hat{\phi}) \delta n_2^{{\rm c}} (\hat{\phi'}) \rangle + \langle \delta n_1^{{\rm c}} (\hat{\phi}) \delta n_2^{\mu}(\hat{\phi'}) \rangle \nonumber \\
          & + & \langle \delta n_1^{\mu}(\hat{\phi}) \delta n_2^{\mu}(\hat{\phi'}) \rangle + \langle \delta n_1^{\mu}(\hat{\phi}) \delta n_2^{{\rm c}} (\hat{\phi'}) \rangle.
\end{eqnarray}
The first term $\langle \delta n_1^{{\rm c}} (\hat{\phi}) \delta n_2^{{\rm c}} (\hat{\phi'}) \rangle$ is due to the intrinsic clustering of the two populations tracing the same large-scale structure,
\begin{equation}
w_{{{\rm cc}}}(\theta) = b_1b_2 \int_0^{\chi_{\rm H}} W_1 W_2 d\chi \int_0^\infty \frac{k}{2\pi} P(\chi, k)J_0(kr\theta) dk,
\end{equation}
where $P(\chi, k)$ is the dark matter power spectrum and $J_0(x) =\sin(x) /x $ is the zeroth-order Bessel function. Note that $W_{\rm cc}$ vanishes if the two populations have disjoint spatial distribution. The second term $\langle \delta n_1^{{\rm c}} (\hat{\phi}) \delta n_2^{\mu}(\hat{\phi'}) \rangle$ is caused by the lensing of the background sources by foreground sources
\begin{eqnarray}
w_{{{\rm fb}}} (\theta) & = & 3 b_1 \Omega_{{\rm m}} (\beta - 1 ) \int_0^{\chi_{{\rm H}}} W_1 \frac{g_2}{a} d\chi \nonumber \\
               & ~ & \int_0^\infty \frac{k}{2\pi} P(\chi, k) J_0(kr\theta) dk.
\end{eqnarray}
The third term $\langle \delta n_1^{\mu}(\hat{\phi}) \delta n_2^{\mu}(\hat{\phi'}) \rangle$ is due to weak lensing by large-scale structure in front of both the foreground and background sources. The last term $\langle \delta n_1^{\mu}(\hat{\phi}) \delta n_2^{{\rm c}} (\hat{\phi'}) \rangle$ represents large-scale structure traced by the background sources lensing the foreground sources which is only present if the two samples have overlapping redshift distributions. The last two terms are negligible. To derive the expected cross-correlations ($w_{{{\rm cc}}}$ and $w_{{{\rm fb}}}$) between our foreground and background samples in Section 4, we use the CAMB software package (Lewis, Challinor \& Lasenby 2000), which is based on CMBFAST (Seljak \& Zaldarriaga 1996), to generate the non-linear matter power spectrum using the fitting formulae of Smith et al. (2003).

\section{Data-sets}

\begin{figure}\centering
\includegraphics[height=3.2in,width=3.5in]{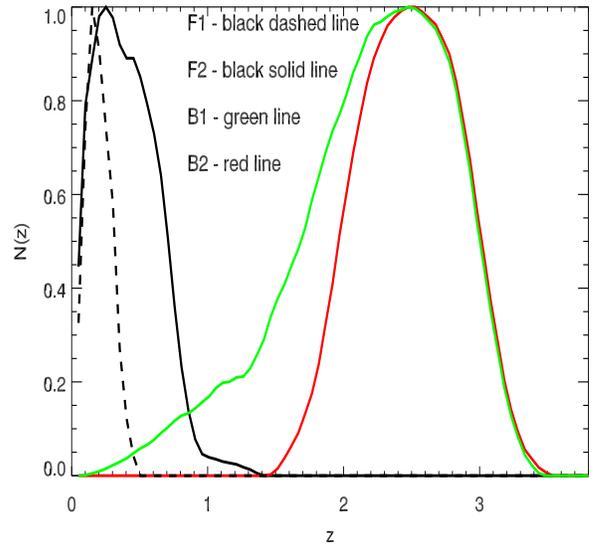}
\caption{Redshift distribution of the foreground and background populations normalised so that the peak of each $N(z)$ is equal to unity. For the foreground sample F1, the $N(z)$ is derived from spectroscopic redshifts. For the foreground sample F2, we have either spectroscopic redshifts or good quality photometric redshifts. The $N(z)$ for the two background samples, B1 and B2, are derived from sub-mm colours using modified black-body spectra.} 
\label{fig:LH_Nz}
\end{figure}

For the first foreground sample, referred to as F1, we select 7,761 sources with $r<19.4$ from the SDSS DR7 in Lockman-SWIRE observed by {\it Herschel}-SPIRE. The star-galaxy separation is done in the same way as in Stoughton et al. (2002). The redshift distribution $N(z)$ of the sample F1 is derived from spectroscopic redshifts obtained in the Galaxy and Mass Assembly (GAMA) survey (Baldry et al. 2010). The median redshift of F1 is $\sim0.2$. The second foreground sample F2 is selected from sources detected by the {\it Spitzer} Infrared Array Camera (IRAC; Fazio et al. 2004) in the SWIRE survey. Full details of the data processing and catalogues can be found in Surace et al. (2005). We {\bf select} 13,888 sources with $S_{3.6} \ge$ 100 $\mu$Jy in the region overlapping with the {\it Herschel}-SPIRE observation in Lockman-SWIRE. The star-galaxy separation is performed in the same way as in Waddington et al. (2007). About 17 percent of the sources in F2 have spectroscopic redshifts and $\sim$80 percent have good quality photometric redshifts with redshift uncertainty $< 0.1$ (Oyaizu et al. 2008; Rowan-Robinson et al. 2008). The median redshift of F2 is $z\sim0.4$.

To construct the background samples in the 13.2 deg$^2$ Lockman-SWIRE field, we use the single-band SPIRE catalogues generated by the SUSSEXtractor source extractor in HIPE (Smith et al. 2010). The cross-match between the 250 and 350 $\mu$m catalogue is done by selecting the brightest 250 $\mu$m source within 12.6 arcsec of a 350 $\mu$m source (FWHM$=$25.2 arcsec at 350 $\mu$m). The flux density at which the integral source counts reach 1 source per 40 beams is 18.7 and 18.4 mJy at 250 and 350 $\mu$m, respectively (Oliver et al. 2010a). The first background sample, B1, comprises sources brighter than 35 mJy at 350 $\mu$m. In total, there are 2,477 / 1,886 such sources in the region that overlaps with F1 / F2. The second background sample B2 includes sources with $S_{350} / S_{250}\gtrsim 0.85$. In total, there are 2,398 / 1,848 such sources in the overlapping region with F1 / F2. About $50\%$ of the sources in B1 are found in B2 as well. Because most of the background sources do not have spectroscopic redshifts, we make use of the sub-mm colours and modified black-body templates to generate qualitative redshift distributions which are consistent with typical model predictions (e.g. Le Borgne et al. 2009; Valiante et al. 2009). The majority of the sources with $S_{350}\gtrsim35$ mJy lie at $1.5\lesssim z \lesssim3$ and peak at $z\sim2$, while most of the sources with $S_{350} / S_{250}\gtrsim 0.85$ lie at $2\lesssim z\lesssim3$ (Amblard et al. 2010; Cooray et al. 2010).

\begin{table}
\caption{Summary of foreground and background samples. The columns are the sample name, the number of sources, the median redshift and the selection criterion. For the two background samples, B1 and B2, we list the number of sources in the overlapping region with F1 and F2, respectively.}\label{table:catalogues}
\begin{tabular}[pos]{llll}
\hline
Sample  & $N_{\rm gal}$      & $\langle z \rangle$    & Selection criterion \\
\hline
F1    & 7,761     & $\sim0.2$  & $r < 19.4$\\
F2    & 13,888    & $\sim0.4$  & $S_{3.6} \ge 100~ \mu $Jy\\
\hline
B1    & 2,477 / 1,886    & $\sim2.0$ & $S_{350} \gtrsim 35 $ mJy\\
B2    & 2,398 / 1,848    & $\sim2.5$ & $S_{350} / S_{250} \gtrsim 0.85$\\
\hline
\end{tabular}
\end{table}

Finally a bright star mask is applied to all samples described above. We follow the procedures in Waddington et al. (2007) and mask a circle around all $K\le12$ point sources in the 2MASS catalogue within a radius $R$ given by $\log$ $R$(arcsec)=3.1 $-$ 0.16$K$. This radius is more conservative compared to the star mask used in the public release of SWIRE catalogues. In Table~\ref{table:catalogues}, we list the number of sources, the median redshift and the selection criteria for the foreground and background samples. Fig.~\ref{fig:LH_Nz} shows the $N(z)$ for each sample. The $N(z)$ of the background is our biggest source of uncertainty. If it is a good approximation, then B2 is almost completely separated from the foreground samples, while B1 has a small overlap with the foreground, in which case $w_{{\rm cc}}$ is non-zero.

\section{Measuring the cross-correlation signal}

The cross-correlation between populations 1 and 2 is the fractional excess in the probability relative to a random distribution (Peebles 1980). We use a modified version of the Landy-Szalay estimator (Landy \& Szalay 1993) to measure the angular cross-correlation function, 
\begin{equation}
w_{\textrm{cross}} (\theta) = \frac{D_1D_2 - D_1R_2 - D_2R_1 + R_1R_2}{R_1R_2},
\end{equation}
where $D_1D_2$, $D_1R_2$, $D_2R_1$ and $R_1R_2$ are the normalised data1-data2, data1-random2, data2-random1 and random1-random2 pair counts in a given separation bin (see Blake et al. 2005 for a disccussion of different estimators of $w_{\textrm{cross}}$). For the foreground samples, we generate random catalogues by distributing sources using a uniform distribution. It is more complicated to generate random catalogues for the background samples. To take into account the noise properties in the sub-mm maps and the angular resolution of SPIRE, we make maps of randomly distributed sources which are processed by the SPIRE Photometer Simulator (SPS; Sibthorpe et al. 2009) for observational programmes exactly the same as the real data. The catalogues extracted from the SPS simulations are then used as random catalogues. To reduce shot-noise in the data-random and random-random pair counts, our random catalogues (after applying the bright star mask) contain roughly 10 times more sources than the real catalogues. We use 40 bootstrap realisations of the foreground and background samples to estimate the errors and covariance matrix.

As described in Section 2, we need the bias factors of the foreground and background samples to calculate the expected clustering-induced and lensing-induced cross-correlations. In the past, sub-mm sources have been shown to cluster strongly (Scott et al. 2002, 2006; Blain et al. 2004; Farrah et al. 2006; Blake et al. 2006; Viero et al. 2009). More recently, the linear bias factor has been measured to be $3.2\pm0.5$ for sources with $S_{350} \gtrsim 30$ mJy and $3.4\pm0.6$ for sources with $S_{350} / S_{250} \gtrsim 0.85$ (Cooray et al. 2010). To derive the bias factors of the foreground samples, we estimate the angular auto-correlation function of F1 and F2, which can be described by a power-law $w_{\textrm{auto}} = A \theta^{-\gamma}$. The amplitude of $w_{\textrm{auto}}$ is related to the correlation length of the spatial correlation function $\xi(r)=(r/r_0)^{-(\gamma+1)}$ (e.g. Efstathiou 1991),
\begin{equation}
A = f r_0^{\gamma} \int \chi^{1-\gamma} \left( N(z) \right)^2 E(z) dz \left(\int N(z) dz \right)^{-2},
\end{equation}
where $f = \sqrt{\pi} \Gamma[(\gamma-1)/2] / \Gamma(\gamma/2)$, $E(z)=( H_0/c ) [\Omega_m(1+z)^3 + \Omega_{\Lambda}]^{1/2}$ and we have assumed constant clustering in comoving units. Finally, we derive the linear bias factor of the foreground using the dark matter correlation function $b = [\xi(r_0) / \xi_{\text{dm}}(r_0)]^{1/2}$. The linear bias factor of F1 and F2 derived in this way is $\sim$1.5 and 1.6 respectively.

\begin{figure*}\centering
\includegraphics[height=4.3in,width=7.2in]{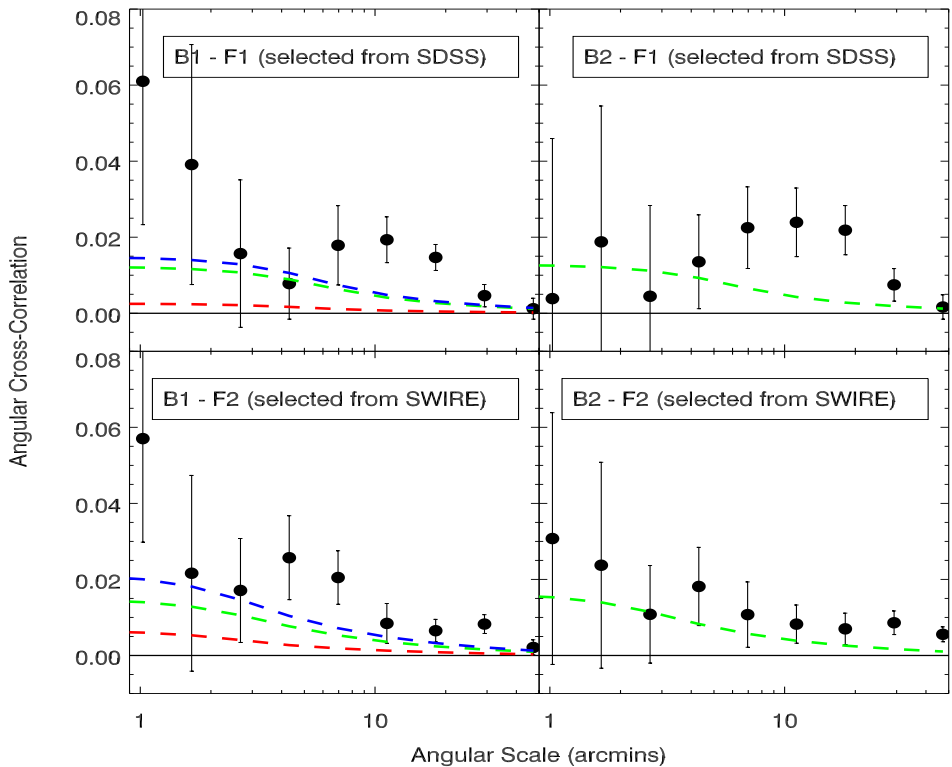}
\caption{The angular cross-correlations between foreground and background populations. The error bars are the rms scatter derived from 40 bootstrap realisations of the real data. In each panel, the red dashed line is the predicted cross-correlation due to mutual clustering $w_{\rm cc}(\theta)$. The green dashed line is the predicted cross-correlation due to lensing $w_{\rm fb}(\theta)$. The blue dashed lines show the sum of $w_{\rm cc}(\theta)$ and $w_{\rm fb}(\theta)$. The black horizontal line denotes the zero level. In the right column, the expected $w_{\rm cc}(\theta)=0$.}
\label{fig:lensing} 
\end{figure*}

The measured angular cross-correlations between the various foreground and background samples are shown in Fig.~\ref{fig:lensing}. A set of logarithmically spaced angular separation bins are used, ranging from $\sim1$ to 50 arcmins. The green dashed line is the expected lensing-induced cross-correlation $w_{\rm fb}(\theta)$, the red dashed line is the expected clustering-induced cross-correlation $w_{\rm cc}(\theta)$ and the blue dashed line is the sum of the two. In the left column of Fig.~\ref{fig:lensing}, the expected clustering-induced cross-correlation $w_{\rm cc}$ is non-zero because the tail of the background $N(z)$ overlaps slightly with that of the foreground $N(z)$. Although $w_{\rm cc}$ is much smaller than $w_{\rm fb}$, we should bear in mind that $w_{\rm cc}$ could be underestimated if a higher than expected fraction of SMGs reside at low redshifts $z\lesssim1$. In the right column of Fig.~\ref{fig:lensing}, the predicted $w_{\rm cc}$ vanishes, as B2 does not overlap with F1 or F2. To assess the significance of the lensing-induced cross-correlation signal, given the covariance matrix obtained from bootstrap realisations, we derive the Bayes factor
\begin{equation}
K=\frac{P(D|M_{\textrm{lensing}})}{P(D|M_{\textrm{null}})},
\end{equation}
where $P(D|M_{\textrm{lensing}})$ is the probability of the data given the lensing model and $P(D|M_{\textrm{null}})$ is the probability of the data assuming there is no cross-correlation. We find that $K=6.3$ for the cross-correlation between F1 and B2 and $K=132.6$ between F2 and B2. On Jeffreys' scale (Jeffreys 1961), $K>3$ means that there is substantial evidence that $M_{\textrm{lensing}}$ is more strongly supported by the data than the null hypothesis and $K>100$ means that there is decisive evidence that $M_{\textrm{lensing}}$ is the favoured model compared to the null. Note that there is almost a factor of two increase in the source density in the foreground sample F2 compared to F1; increasing the number of tracers of the foreground structure increases the strength of the lensing signal.

\section{The effect of weak lensing on the number count of sub-mm sources}

The effect of lensing on the number count of the sub-mm sources is expressed in equation (2), under the assumption that the lens plane is at a much lower redshift than the source plane. The power-law slope of the intrinsic / unlensed number count $N_{{\rm u}}(S)$ is not affected because the lensing magnification $\mu$ is independent of the flux density. However, the overall normalisation of the number count can be modified by a factor of $\mu^{\beta-1}$, where $\mu=1+\delta \mu=1+2\kappa$ in the weak lensing limit. Weak lensing by large-scale structure causes $\delta \mu$ to follow a Gaussian function with mean magnification $\langle \delta\mu \rangle=0$ and its dispersion $\sigma_{\mu}$ dependent on the redshift of the sub-mm population (BS01). Therefore, when averaged over a statistically representative area, the effect of weak lensing on the number count should be negligible.

The effect of weak lensing on the local number density of the sub-mm sources along a certain direction can be estimated from the measured cross-correlation between the foreground and the background populations. In the right panel of Fig. 2 where the measured signal is expected to be due to lensing only, we can see that the probability of finding a background sub-mm source close to a foreground galaxy is increased by a few percent above random on angular scales between $\sim1$ and 50 arcmin. Therefore, the lensing induced change in the number density along a certain direction is expected to be at the level of a few percent.

We can also estimate the effect of lensing on the local number density through the auto-correlation function of the background sub-mm sources, $w_{\rm auto}(\theta)=\langle \delta n (\hat{ \phi }) \delta n( \hat{\phi'} ) \rangle $. Using equation (4), we can decompose $w_{\rm auto}(\theta)$ into three components, $\langle \delta n^c(\hat{\phi}) \delta n^c(\hat{\phi'}) \rangle$, $\langle \delta n^c(\hat{\phi}) \delta n^{\mu}(\hat{\phi'})\rangle + \langle \delta n^c(\hat{\phi'}) \delta n^{\mu}(\hat{\phi})\rangle$ and $\langle \delta n^{\mu}(\hat{\phi}) \delta n^{\mu}(\hat{\phi'}) \rangle$, which represent the galaxy-galaxy, galaxy-lensing and lensing-lensing correlation functions respectively. The lensing-lensing term is given by (Moessner \& Jain 1998)
\begin{eqnarray}
w_{{{\rm auto}}}^{{\rm lensing-lensing}} (\theta) & = & [3 \Omega_{{\rm m}} (\beta - 1 )]^2 \int_0^{\chi_{{\rm H}}} (g_2/a)^2 d\chi \nonumber \\
               & ~ & \int_0^\infty \frac{k}{2\pi} P(\chi, k) J_0(kr\theta) dk.
\end{eqnarray}
At zero lag, $w(0)_{\rm auto}^{\rm lensing-lensing}=\langle (\delta n^{\mu})^2 \rangle$ is the variance of the number density fluctuatuation due to lensing and thus the rms fluctuation is $\delta n^{\mu} = (w(0)_{\rm auto}^{\rm lensing-lensing})^{1/2}$ which is at a few percent level.

\section{Discussions and conclusions}
\label{discussions and conclusions}

The unusually steep number count in the bright sub-mm regime leads to an enhanced cross-correlation signal that is due to weak gravitational lensing. In this paper, we have measured the angular cross-correlations between sub-mm sources detected by {\it Herschel}-SPIRE in Lockman-SWIRE and foreground sources selected in the optical or near-infrared. We have also derived theoretical expectations of the weak lensing-induced cross-correlation $w_{\rm fb}$ and the clustering-induced cross-correlation $w_{\rm cc}$ which are in good agreement with our measurements. We find clear evidence for a lensing-induced cross-correlation between sub-mm sources at high redshifts and galaxies at low redshifts.

The redshift distribution of the sub-mm sources is the biggest source of uncertainty in our analysis because most of the sources do not have spectroscopic redshifts. In principle, the clustering-induced cross-correlation $w_{\rm cc}$ could contaminate the lensing-induced cross-correlation $w_{\rm fb}$ if a higher than expected fraction of sub-mm sources reside in the low-redshift Universe. As the amplitude of $w_{\rm fb}$ is mainly sensitive to the mean redshift of the background population rather than the exact shape of the $N(z)$ (M\'{e}nard \& Bartelmann 2002), we have carried out a simple calculation of the expected $w_{\rm fb}$ and $w_{\rm cc}$ amplitude by varying the mean redshift $\langle z \rangle$ (from 0.3 to 4.0) and the width $\sigma_z$ (from 0.2 to 2.5), assuming the $N(z)$ of the sub-mm sources can be approximated by a Gaussian function. In all cases, to reproduce the measured cross-correlation signal, $w_{\rm cc}$ is at most comparable to $w_{\rm fb}$ when $\langle z \rangle \sim 3.5, \sigma_z \sim 1.5$, $\langle z \rangle \sim 2.5, \sigma_z \sim 1.0$ or $\langle z \rangle \sim 1.5, \sigma_z \sim 0.5$. So the detection of the weak lensing-induced cross-correlation should be robust. It should be possible to acurately determine $N(z)$ in the future when the infrared spectral energy distributions are well understood and/or more spectroscopic redshifts are acquired for sub-mm sources.

Limitations in our modelling of the cross-correlation include: using a scale- and time-independent bias factor for the galaxy-dark matter power spectrum; assuming a linearised magnification; and adopting a constant power-law number count slope independent of flux. While for this first study a simple model is adequate given the large error bars, an approach such as the halo model to describe the galaxy-dark matter power spectrum can be utilised in the future when additional data warrant an improved description (e.g. Jain et al. 2003). The expected increase in area covered by {\it Herschel}-SPIRE will allow the detection of cosmic magnification presented in this paper to be improved and be used to constrain cosmological parameters and galaxy bias.

\section*{ACKNOWLEDGEMENTS}
LW is suppported by UK's Science and Technology Facilities Council grant ST/F002858/1. LW thanks Antony Lewis for helpful discussions. We thank the GAMA team for providing the redshift distribution of F1. The data presented in this paper will be released through the Herschel database in Marseille HeDaM (hedam.oamp.fr/herMES). SPIRE has been developed by a consortium of institutes led by Cardiff Univ. (UK) and including Univ. Lethbridge (Canada); NAOC (China); CEA, LAM
(France); IFSI, Univ. Padua (Italy); IAC (Spain); Stockholm Observatory
(Sweden); Imperial College London, RAL, UCL-MSSL, UKATC, Univ. Sussex
(UK); Caltech, JPL, NHSC, Univ. Colorado (USA). This development has been
supported by national funding agencies: CSA (Canada); NAOC (China); CEA,
CNES, CNRS (France); ASI (Italy); MCINN (Spain); SNSB (Sweden); STFC (UK);
and NASA (USA).

\end{document}